\documentclass[prl,twocolumn,showpacs]{revtex4}

\usepackage{graphicx}
\usepackage{amsmath}

\usepackage{amsmath}
\usepackage{amsfonts}

\newcommand{\Z}{\mathbb{Z}}

\def\be{\begin{equation}}
\def\ee{\end{equation}}
\def\ba{\begin{eqnarray}}
\def\ea{\end{eqnarray}}

\begin{document}

\title{Exact solution of the anisotropic special transition in the $\mathcal{O}(n)$ model in 2D}

\author{J\'er\^ome Dubail$^{1}$, Jesper Lykke Jacobsen$^{2,1}$ and Hubert Saleur$^{1,3}$}
\affiliation{${}^1$Institut de Physique Th\'eorique, CEA Saclay,
91191 Gif Sur Yvette, France}
\affiliation{${}^2$LPTENS, 24 rue Lhomond, 75231 Paris, France}
\affiliation{${}^3$Department of Physics,
University of Southern California, Los Angeles, CA 90089-0484}

\date{\today}

\begin{abstract}

  The effect of surface exchange anisotropies is known to play a
  important role in magnetic critical and multicritical behavior at
  surfaces. We give an exact analysis of this problem in $d=2$ for the
  ${\cal O}(n)$ model by using Coulomb gas, conformal invariance and
  integrability techniques. We obtain the full set of critical
  exponents at the anisotropic special transition---where the symmetry
  on the boundary is broken down to ${\cal O}(n_1)\times {\cal
    O}(n-n_1)$---as a function of $n_1$. We also obtain the full phase
  diagram and crossover exponents. Crucial in this analysis is a new
  solution of the boundary Yang-Baxter equations for loop models. The
 appearance  of the generalization of Schramm-Loewner
 Evolution $SLE_{\kappa,\rho}$ is also discussed.

\end{abstract}



\maketitle


The study of boundary critical phenomena is relevant to a very large
number of physics problems. These include, on the condensed matter
side, the critical behavior of magnets and alloys with free surfaces
\cite{review1}, adsorption of fluids or polymers on walls and
interfaces \cite{review2}, but also---through a series of by now well
known mappings---the Kondo and other effects in quantum impurity
problems \cite{review3,review4}. On the high-energy physics side,
apart from the old problem of studying field theories on manifolds
with boundaries, the more recent developments inspired by string
theory have put boundary effects squarely in the limelight. This
includes the physics of D-branes \cite{review5}, and of course the
celebrated AdS/CFT conjecture \cite{review6}. Boundary effects are
also central to many recent developments in the study of geometrical
critical phenomena and the Schramm Loewner Evolution approach
\cite{review7}.

Going back to the critical behavior of magnets, surface effects can
give rise to a bewildering array of physical effects \cite{review8},
especially when combined with finite size effects \cite{review9}. The
experimental activity \cite{exp1,exp2,exp3} has been
steady.

In interpreting experimental results, a natural question concerns the
effects of surface anisotropy \cite{exp4}. While it was quickly
understood that such effects are irrelevant near the ordinary
transition in the $n$-vector model \cite{Diehl1}, further study
\cite{Diehl2} showed that they are relevant near special transitions,
and that they lead, in high enough dimension, to the emergence of new
``anisotropic special'' transitions. The corresponding family of new
boundary critical points was extensively studied in $d=4-\epsilon$
(see \cite{Diehl3} for recent work), and in the $1/n$ expansion
\cite{Ohno}. The presence of anisotropy on the boundary is
particularly interesting in $d=3$ since, while only ordinary
transitions are observable for isotropic systems with continuous
symmetry, the presence of an easy axis on the boundary allows for an
(anisotropic) special transition, which has been studied to second
order in $\epsilon$ \cite{Diehl2}.

We present in this Letter a complete solution of the problem in $d=2$,
in the context of a geometrical reformulation of the $n$-vector model
as a loop model \cite{Nienhuis}. Even though the ``surface'' is here
one-dimensional---and so strictly speaking cannot order for integer
$n$---this reformulation in fact exhibits all the physics of the
transitions in higher dimension. In particular, we fully recover the
phase diagram conjectured in \cite{Diehl2}.  Moreover, the loop
formulation permits to treat $n$ as a real variable, and the limits $n
\to 0$ and $n \to 1$ give access to physical results for polymers and
the Ising model.

The case $d=2$ is interesting for other reasons as well. One concerns
the classification of all possible conformal boundary conditions for
``non-minimal'' conformal field theories (CFTs)---such as the loop
models \cite{JS}---with potential applications to string theory or 2D
quantum gravity \cite{Kostov}. Another has to do with the general
program of understanding all critical exponents in geometrical models,
as well as their relations with (variants of) the SLE
formalism \cite{Cardy}.

\paragraph{Loop model.}

The classical ${\cal O}(n)$ model in $d$ dimensions is defined, initially, by placing on each
lattice site $i$ a vector spin $\vec{S}_i$ with components $S_i^\mu$
for $\mu=1,\ldots,n$.  Along each link $(ij)$, neighboring spins
interact through the Boltzmann weight $\exp(x \vec{S}_i \cdot
\vec{S}_j)$. At high temperatures (low $x$) this can be replaced by
$1+x \vec{S}_i \cdot \vec{S}_j$. This replacement does not modify the
long-distance behavior, up to and including the critical point $x_{\rm
  c}$. The partition function then reads
\be
 Z = \mathrm{Tr} \prod_{\langle ij \rangle}
 \left( 1+ x \vec{S}_i \cdot \vec{S}_j \right) \,,
 \label{Z1}
\ee
where $\langle ij \rangle$ denotes the set of nearest neighbors.  The
trace over spin configurations can be normalized so that ${\rm Tr} \
S_i^\mu S_j^\nu = \delta_{ij} \delta_{\mu \nu}$.

Expand now the product in (\ref{Z1}), and draw a monomer on link
$(ij)$ when the term $x S_i^\mu S_j^\mu$ is taken. The trace of terms
containing an odd power of any $S_i^\mu$ vanishes by the symmetry
$\vec{S}_i \to -\vec{S}_i$. Specializing to a trivalent lattice,
non-vanishing terms can then only contain $S_i^\mu$ to the power $0$
or $2$. The monomers drawn hence form configurations ${\cal C}$ of
non-intersecting loops. Summing over components then yields
\be
 Z = \sum_{\cal C} x^X n^N \,,
 \label{Z2}
\ee
where $X$ (resp.\ $N$) is the number of monomers (resp.\ loops) in
${\cal C}$. In this formulation, $n$ can be considered a continuous
variable. When $x \nearrow x_{\rm c}$, the average
length of a loop passing through the origin diverges.

\paragraph{Surface critical behavior.}

\begin{figure}[h]
\center
\includegraphics[width=0.32\textwidth]{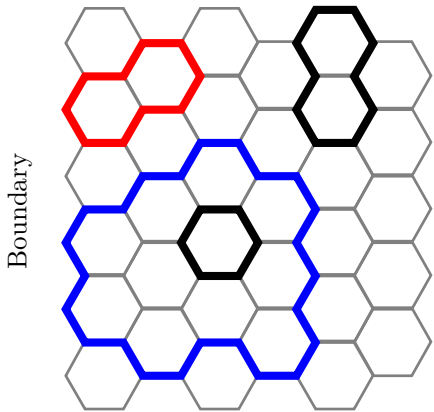}
\caption{${\cal O}(n)$ loop model with a boundary. Boundary loops pass through
one or more boundary sites, and can be of type 1 (drawn blue) or 2 (red). Bulk loops are drawn black.}
\label{fig:blob1b}
\end{figure}

When a boundary is present (see Fig.~\ref{fig:blob1b}), boundary sites have fewer
neighbors than bulk sites, so we can expect boundary spins to be disordered
at and slightly above $x_{\rm c}$, where bulk spins start ordering. In
the loop picture, this means that boundary monomers are less probable
than bulk monomers, and in the continuum limit loops will avoid the boundary.
This is the ordinary surface transition {\em Ord}.

Consider  now the model
%
 $Z = \sum_{\cal C} x^X w^W n^N$
%
with an extra weight $w$ each time a loop passes through a boundary
site (see Fig.~\ref{fig:blob1b}). When $w > 1$, loops are attracted
to the boundary. At a critical value $w_{\rm c}$, this attraction
precisely compensates the lower number
of neighbors, so that bulk and boundary spins order simultaneously.
This is the special surface transition {\em Sp}.
All other values of $w$ flow towards $1$ or $\infty$, the latter being
the extraordinary transition {\em Ex} in which a single loop occupies
the whole boundary. The set of boundary monomers has a non-trivial
fractal dimension, $0 < d_{\rm f} < d-1$, only at {\em Sp}.

For $d=2$ the above references to spin ordering do not make sense due
to the Mermin-Wagner theorem. The transitions {\em Ord} and {\em Sp}
nevertheless exist in the {\em loop} model. On the honeycomb lattice,
for $-2 < n \leq 2$, $x_{\rm c} = (2+\sqrt{2-n})^{-1/2}$
\cite{Nienhuis90} and  $w_{\rm c} = 1 + 2/\sqrt{2-n}$ \cite{Batchelor95}.

\paragraph{Surface anisotropy.}

Motivated by experiments \cite{exp4} and theoretical developments for
$d>2$ \cite{Diehl1,Diehl2,Diehl3,Ohno}, we now allow for anisotropy 
in the boundary interaction, breaking the symmetry down to 
${\cal O}(n_1) \times {\cal O}(n_2)$. The effect in the loop model formulation is immediate.
 We call type 1 (drawn blue
in Fig.~\ref{fig:blob1b}) a loop for which $\mu=1,\ldots,n_1$, and
type 2 (red) a loop with $\mu=n_1+1,\ldots,n$. For boundary loops, the
sum over $\mu$ is done separately for each loop type, whereas for bulk
loops the summation is complete as before. The fugacity of boundary
loops of type 1 (resp.\ 2) is thus $n_1$ (resp.\ $n_2 = n-n_1$). This
leads to
\be
 Z = \sum_{\cal C} x^X w_1^{W_1} w_2^{W_2} n^N n_1^{N_1} n_2^{N_2} \,,
 \label{Z4}
\ee
where now $N$ is the number of {\em bulk} loops only, and we have
introduced type-dependent weights $w_1$ and $w_2$ for each time a loop
passes through a boundary site. [The total weight in Fig.~\ref{fig:blob1b}
is the product of that of the blue (resp.\ red) boundary loop, $x^{24} w_1^2 n_1$ (resp.\
$x^{10} w_2 n_2$), and that of the bulk loops (black), $x^6 n \times x^{10} n$.] We
henceforth consider $n_1$ a continuous variable,
with $0 \le n_1 \le n$.

\paragraph{Integrability.}

As in previous work \cite{Nienhuis90,Batchelor95}, our exact results
for the honeycomb loop model are derived from the integrability of a
related model on the square lattice.  It is defined by the vertex
weights
\begin{center}
\includegraphics[width=0.47\textwidth]{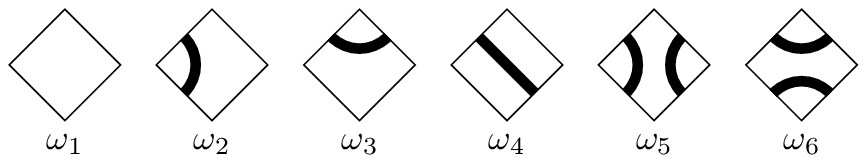}
\end{center}
where vertices related by horizontal and vertical reflection have been
drawn only once. This model is integrable in the bulk---i.e., the weights satisfy
the Yang-Baxter equations with spectral parameter $u$---when
\cite{Nienhuis90}
\be
 \label{YBweights}
 \begin{array}{rcl}
 \omega_1 &=& \sin 2\psi \sin 3\psi + \sin u \sin (3\psi-u)\\
 \omega_2 &=& \sin 2\psi \sin (3\psi-u) \\
 \omega_3 &=& \sin 2\psi \sin u \\
 \omega_4 &=& \sin u \sin (3\psi-u) \\
 \omega_5 &=& \sin (2\psi-u) \sin (3\psi-u) \\
 \omega_6 &=& - \sin u \sin (\psi-u) \,,
\end{array}
\ee
and $\psi$ parametrizes $n = -2 \cos 4 \psi$.
Boundary interactions are introduced as follows
\begin{center}
\includegraphics[width=0.24\textwidth]{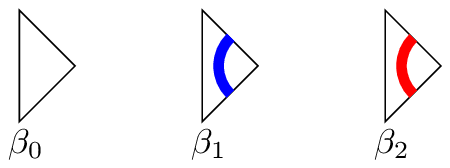}
\end{center}
The following integrable weights---i.e., satisfying the boundary
Yang-Baxter equation---describe the well-known isotropic transitions
\cite{Batchelor95}
\be
 \begin{array}{lll}
 \mbox{{\em Ord}:} \ &
 \beta_0 = \sin \left( \frac{3}{2} \psi + u \right), &
 \beta_1 = \beta_2 = \sin \left( \frac{3}{2} \psi - u \right) \\
 \mbox{{\em Sp}:} &
 \beta_0 = \cos \left( \frac{3}{2} \psi + u \right), &
 \beta_1 = \beta_2 = \cos \left( \frac{3}{2} \psi - u \right) 
\end{array}
\nonumber
\ee
We have generalized this result by finding a new  anisotropic solution $AS_1$, with
$n_1 = -\sin \left( 4(\eta-1)\psi \right)/\sin\left( 4 \eta \psi\right)$ parametrized by $\eta$:
\be
 \label{Sklyanin}
 \begin{array}{rcl}
 \beta_0 &=&  \sin \left((2 \eta +\frac{1}{2}) \psi-u\right)
 \sin \left((2 \eta - \frac{1}{2})\psi +u\right) \\
 \beta_1 &=& \sin \left((2\eta+\frac{1}{2}) \psi+u\right)
 \sin \left((2 \eta - \frac{1}{2})\psi +u\right) \\
 \beta_2 &=& \sin \left((2 \eta +\frac{1}{2}) \psi-u\right) \sin \left((2 \eta - \frac{1}{
2})\psi -u\right) \,.
 \end{array}
\ee
Another solution $AS_2$ arises from the duality transformation that
exchanges the two loop types, i.e., $n_1 \to n - n_1$.

When $u=\psi$ we have $\omega_6=0$ in (\ref{YBweights}), and so each
vertex can be pulled apart horizontally so as to form a pair of honeycomb 
vertices \big(\includegraphics[width=0.07\textwidth]{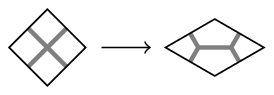}\big). The weights in (\ref{Z4}) read $w_{1,2}=\sqrt{ \beta_{1,2} / (x \beta_0) }$, evaluated \cite{Batchelor95,inprep} in $u/2$:
\be
 w_{1,2} = 1 + \frac{\sqrt{2-n}}{2} \pm
 \frac{n_1-\frac{n}{2}+\sqrt{1-n_1(n-n_1)}}
 {\sqrt{2-n}} \,.
 \label{w12}
\ee

\paragraph{Continuum limit.}

The long distance behavior at the anisotropic special point $AS_1$
could be inferred by setting up the Bethe Ansatz equations
corresponding to (\ref{Sklyanin}). An alternative and easier route
is to use the Coulomb gas approach to CFT \cite{Nienhuis}.

Consider first the bulk theory. Orient each loop independently, and
attribute a weight $e^{i\gamma \alpha/(2\pi)}$ when a loop turns an
angle $\alpha$.  Summing over orientations, this gives
$n=2\cos\gamma$. A height field $h$ is defined by viewing oriented
loops as it level lines, across which $h \to h \pm \pi$. This is
well-known \cite{Nienhuis} to renormalize towards a Gaussian free
field with action
\be
 S = \frac{g}{4\pi} \int (\partial h)^2 \, {\rm d}^2 x \,.
 \label{Gaussian}
\ee
The coupling constant $g$ is fixed by requiring that the operator
$\cos(2h)$ conjugate to the discretization of $h$ (and hence to
the weight $n$) be strictly marginal. This gives $g=1+\gamma/\pi$.
The central charge $c$ and the Kac critical exponents $h_{a,b}$ of
primary operators $\Phi_{a,b}$ are parametrized by $a,b$ and $g$:
\be
 c = 1 - 6 \frac{(g-1)^2}{g} \,, \quad
 h_{a,b} = \frac{(g a - b)^2 -(g-1)^2}{4 g} \,.
 \label{Kac}
\ee

It is convenient to define the boundary theory on an $L \times T$
annulus with period $T$. Its left rim roles as the boundary
(cf.~Fig.~\ref{fig:blob1b}). Type 1 loops have weight $n_1 =
\sin\left((r+1)\gamma\right)/\sin(r \gamma)$, with $r \in \left( 0,\pi/\gamma \right)$
a new parameter. The operators that constrain a boundary monomer to be of
type 1 or 2 are orthogonal projectors. We can write, e.g., the type 1
projector in the basis of oriented loops:
\begin{center}
\includegraphics[width=0.47\textwidth]{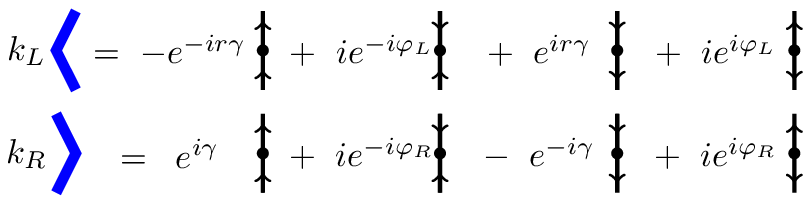}
\end{center}
where the top (resp.\ bottom) line pertains to the left (resp.\
right) rim of the annulus, and $k_{L} = 2 i \sin r \gamma$, $k_{R} = 2 i \sin \gamma$.
Loops touching only the right rim are bulk loops and have a weight
$n$. Requiring weight $n_1$ for loops touching both
rims fixes $\varphi_L-\varphi_R=r \gamma$. Note that our interaction does not conserve the arrow current. 

We assume that the diagonal part induces a flow towards fixed boundary conditions for the dual field.
 We thus end up with a free field with Neumann boundary conditions on both rims, and
additional weights $e^{\pm i r \gamma}$ for each of the $p$ pairs of
oriented half-loops going from one rim to the other. This amounts to a
height defect $\Delta h = 2 \pi p$ when going around the periodic
direction, $y \to y + T$, of the annulus. This can be gauged away by
writing $h(x,y) = \tilde{h}(x,y) + 2\pi p y / T$, where now
$\tilde{h}$ is periodic. The second term gives rise to
\be
 \sum_{p \in \Z} e^{i p r\gamma} e^{-\left( \frac{g}{4\pi} \right)
 p^2 \left( \frac{2\pi}{T} \right)^2 L T} \propto
 \sum_{n \in \Z} e^{- \left( \frac{\pi T}{4g L} \right)
 \left( \frac{r \gamma}{\pi} -2n \right)^2} \,,
 \nonumber
\ee
where we used (\ref{Gaussian}) and a Poisson resummation.
Integration over the first term gives $q^{-1/24} / P(q)$, with modular
parameter $q=e^{- \pi T/L}$ and $P(q)=\prod_{k \geq 1}
\left(1-q^k\right)$. Using now (\ref{Kac}) gives the exact
continuum limit partition function in the sector with zero
non-contractible loops:
\be
 Z_0(q) = \frac{q^{-c/24}}{P(q)} \sum_{n \in \Z} q^{h_{r,r-2 n}} \,,
 \label{Z_0}
\ee
where the easy limit $T \gg L$ has been used to fix the normalization.
The complete spectrum of critical exponents can be read off from (\ref{Z_0}).

To be precise, (\ref{Z_0}) is valid at the transition $AS_2$. This can
be seen from the limit $n_1 \to n$, under which the leading exponent
$h_{r,r}$ in (\ref{Z_0}) vanishes.  This fits with $AS_2 \to Ord$,
while $AS_1 \to Sp$. However, results for $AS_1$ can easily be
obtained by applying the duality transformation to (\ref{Z_0}).
We conclude that the boundary condition changing (BCC) operators are
$\Phi_{r,r+1}$ for $(AS_1/Ord)$ and $\Phi_{r,r}$ for $(AS_2/Ord)$.

\paragraph{Fractal dimensions.}

A non-contractible loop on the annulus is generated by the operator
$\Phi_{2,1}$ \cite{review7}. Conformal fusing with the BCC operator
gives $\Phi_{r,r+1} \otimes \Phi_{2,1} = \Phi_{r+1,r+1} \oplus
\Phi_{r-1,r+1}$ for $AS_1$. To interpret this, note that the first
term on the right-hand side is dominant, and since $w_1 > w_2$ in
(\ref{w12}) this must correspond to the insertion of a type 1 (blue)
non-contractible loop. The second term thus produces a type 2 (red)
loop.

The fractal dimension $d_f^{(1)}$ of the set of type 1 boundary
monomers is conjugate to the operator inserting two blue loop strands
at the boundary. This is obtained by fusing $\Phi_{r+1,r+1}$ with
itself, giving the principal contribution $\Phi_{2r+1,2r+1}$.
Therefore
\be
\label{eq:fractal}
 d_f^{(1)} = 1-h_{2r+1,2r+1}=1 - r(r+1) \frac{(g-1)^2}{g}
\ee
at $AS_1$. This is non-trivial ($0 < d_f < 1$), just as the
result \cite{Batchelor95} $d_f=1-h_{3,3}=1 - 2 (g-1)^2/g$ at the
special transition. We find similarly that $d_f^{(2)} < 0$ at $AS_1$.

These results imply the physical interpretation of $AS_1$: type 1
loops are critically attracted towards the boundary, and type 2 loops
retract from it. In other words, type 1 (resp.\ type 2) loops stand at
an anisotropic special (resp.\ ordinary) transition.

\paragraph{Phase diagram.}

\begin{figure}[h]
\centering
\includegraphics[width=0.45\textwidth]{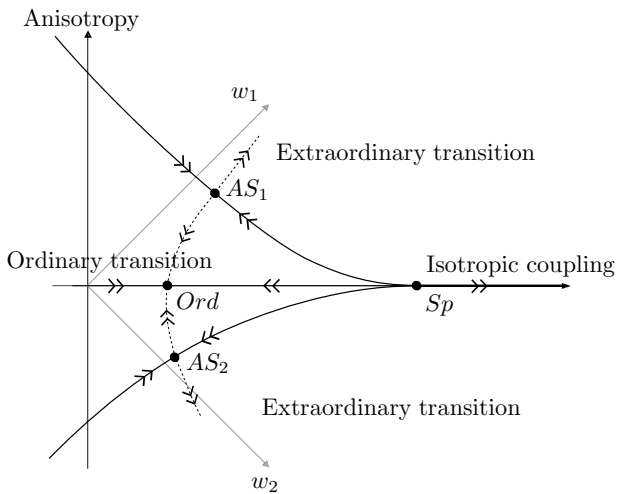}
\caption{Generic phase diagram for $0 < n_1< n$ in the rotated $(w_1,w_2)$
plane.}
\label{fig:phase}
\end{figure}

We are now ready to propose the phase diagram of the model (\ref{Z4}),
for $x=x_{\rm c}$ and in the regime $0 < n_1 < n$. See
Fig.~\ref{fig:phase}.  The fixed points (depicted as solid circles) are
conformally invariant boundary conditions, and the double arrows
represent flows under the boundary renormalization group (RG).

Surface anisotropy must be relevant (resp.\ irrelevant) at the special
(resp.\ ordinary) isotropic transition, since the loops see (resp.\ do
not see) the boundary in that case. Only in the former case can we
expect $n_1$ to change the critical behavior of boundary loops. This
agrees physically with the conclusions of \cite{Diehl1,Diehl2} for
$d>2$.

More precise evidence for these flows can be found by evaluating the
boundary entropies $S_{\rm b} = -\log g_{\rm b}$ for the various boundary
conditions. By the so-called $g$-theorem \cite{AffleckLudwig}, $S_{\rm b}$
increases under the boundary RG flows: the flow is from a less
stable to a more stable boundary condition. $g_{AS_1}$ and $g_{AS_2}$ are related by $n_1 \rightarrow n-n_1$. We find \cite{inprep}
\be
 g_{AS_1}=\left( \frac{2}{g} \right)^{1/4} \frac{\sin ((r+1) \gamma/g)}{\sin r \gamma} \left( \frac{\sin \gamma}{\sin (\gamma/g)} \right)^{1/2}
\ee
to be compared with
$g_{Ord} = (2/g)^{1/4} \big( \sin(\gamma/g) / \sin \gamma \big)^{1/2}$ and
$g_{Sp} = (2/g)^{1/4} \sin(2\gamma/g) /
\big(\sin\gamma \sin(\gamma/g)\big)^{1/2}$.
We have thus $S_{Ord} > S_{AS_{1,2}} > S_{Sp}$, as is consistent with the
flows of Fig.~\ref{fig:phase}. Note that it is possible to flow to $AS_1$
by tuning only $w_1$ and not $w_2$, in agreement with the interpretation
that only type 1 loops stand at $AS_1$.

Near the point $Sp$ we have $w_1-w_2 \sim (w_{\rm c}-w)^{1/\phi}$. Identifying
the operators perturbing in the isotropic and anisotropic directions, and
using standard scaling arguments, gives the cross-over exponent
$\phi = (1-h_{1,3})/(1-h_{3,3})$. The fact that $\phi<1$ for $0 <n \le 2$
implies the cusp-like shape of the phase diagram near $Sp$. This feature
is also present \cite{Diehl2} in $d>2$.

\paragraph{Physical realization of $SLE_{\kappa,\rho}$.}
Schramm-Loewner Evolutions have a natural generalization with a distinguished boundary point, leading to the two-parameter family of processes $SLE_{\kappa,\rho}$ \cite{review7,Cardy}. Its physical relevance has however remained unclear so far.

Our geometric formulation provides a lattice object whose scaling limit must be described by such a process. Indeed, according to the SLE/CFT correspondance \cite{review7}, type $1$ loops (blue) at $AS_1$ correspond to $\kappa = 4/g$ and $\rho = \kappa \left( h_{r+1,r+1}-h_{r,r+1} -h_{2,1} \right) =r(4-\kappa)/2-2$. The fractal dimension of the intersection of the $SLE_{\kappa,\rho}$ trace with the real axis is ($\ref{eq:fractal}$): $d_f^{(1)}=\left(1 + \rho/4 \right) \left(2 -8/\kappa - 4 \rho/\kappa \right)$.

\smallskip

\paragraph{Acknowledgments.}

We thank I. Kostov and D. Bernard for helpful exchanges.  This work was supported by
the European Community Network ENRAGE (grant MRTN-CT-2004-005616) and
by the Agence Nationale de la Recherche (grant ANR-06-BLAN-0124-03).

\end{document}